\documentclass{aa}

\usepackage{graphicx}
\usepackage{txfonts}
\usepackage{natbib}
\bibpunct{(}{)}{;}{a}{}{,}

\begin{document}

\title{Planets around evolved intermediate-mass stars}

\subtitle{I. Two substellar companions in the open clusters NGC~2423 and NGC~4349\thanks{Based on observations made with the ESO 3.6m-telescope at La Silla Observatory under program IDs 075.C-0140, 076.C-0429, 077.C-0088 and 078.C-0133.}}

\author{C. Lovis
\and M. Mayor
}

\offprints{C. Lovis}

\institute{Geneva Observatory, University of Geneva, 51 ch. des Maillettes, 1290 Sauverny, Switzerland\\
\email{christophe.lovis@obs.unige.ch}
}

\date{Received 28 February 2007 / Accepted xx xxxx 2007}

\abstract
{Many efforts are being made to characterize extrasolar planetary systems and unveil the fundamental mechanisms of planet formation. An important aspect of the problem, which remains largely unknown, is to understand how the planet formation process depends on the mass of the parent star. In particular, as most planets discovered to date orbit a solar-mass primary, little is known about planet formation around more massive stars.}
{To investigate this point, we present first results from a radial velocity planet search around red giants in the clump of intermediate-age open clusters. We choose clusters harbouring red giants with masses between 1.5 and 4~$M_{\sun}$, using the well-known cluster parameters to accurately determine the stellar masses. We are therefore exploring a poorly-known domain of primary masses, which will bring new insights into the properties of extrasolar planetary systems.}
{We are following a sample of about 115 red giants with the Coralie and HARPS spectrographs to obtain high-precision radial velocity (RV) measurements and detect giant planets around these stars. We use bisector and activity index diagnostics to distinguish between planetary-induced RV variations and stellar photospheric jitter.}
{We present the discoveries of a giant planet and a brown dwarf in the open clusters NGC~2423 and NGC~4349, orbiting the 2.4~$M_{\sun}$-star NGC2423~No3 (TYC~5409-2156-1) and the 3.9~$M_{\sun}$-star NGC4349~No127 (TYC~8975-2606-1). These low-mass companions have orbital periods of 714 and 678~days and minimum masses of 10.6 and 19.8~$M_\mathrm{Jup}$, respectively. Combined with the other known planetary systems, these detections indicate that the frequency of massive planets is higher around intermediate-mass stars, and therefore probably scales with the mass of the protoplanetary disk.}
{}


\maketitle

\section{Introduction}

More than 200~extrasolar planets have been discovered over the past decade, revealing a completely unexpected diversity in the properties of planetary systems \citep{udry07}. Models of planet formation had to be largely revised to explain a number of characteristics derived from the observations such as the planet-metallicity correlation, orbital migration, and the occurrence of high eccentricities. Increasingly complex models including Monte Carlo simulations are being developed by different groups \citep[e.g.][]{ida04,alibert05} with the goal of comparing theoretical predictions to the observed distribution of extrasolar planets.

Besides metallicity, protoplanetary disk masses and surface densities are other important factors suspected to strongly influence the formation of giant planets. They are thought to be dependent on stellar mass, in the sense that more massive stars will have more massive disks and higher surface densities \citep[see][]{ida05}, although this point needs confirmation. The exact impact of this on planet formation is presently poorly known theoretically but the favoured core-accretion scenario predicts that massive giant planets should form more frequently around solar-mass stars than around low-mass stars \citep{laughlin04,ida05}. This is also supported by observational data showing that giant planets are much rarer around M~dwarfs than around solar-type stars \citep{bonfils05,endl06,butler06}.

In this context, it is of great interest to investigate the frequency of giant planets around stars significantly more massive than the Sun. Unfortunately, it is impossible to apply usual high-precision Doppler techniques to stars with spectral types earlier than late-F due to the increase in rotational velocities and the low number of spectral lines in these stars. A few attempts have been made, however, to adapt the technique to A--F stars \citep{galland05}, unveiling a brown dwarf candidate around a 1.7~$M_{\sun}$ primary \citep{galland06}.

Another promising approach is to study intermediate-mass stars in a more evolved stage, i.e. in the red giant phase, where rotational velocities are low and many spectral lines are available for Doppler shift measurements. However, great care has to be taken when interpreting RV variations in red giants due to the presence of intrinsic stellar jitter. Precise RV measurements have already been reported for many red giants in the solar neighbourhood, and a few planet candidates have been discovered around such stars \citep{frink02,hatzes03,hatzes05,hatzes06,setiawan03,setiawan05,sato03}. These detections demonstrate that in these cases stellar jitter does not hide the RV signal of giant planets. It appears that a significant fraction of red giants are indeed suitable for planet search, provided they are chosen not too evolved or too cool \citep{sato05}.

In the context of planet formation around intermediate-mass stars, it is crucial to have a precise knowledge of primary masses. This is unfortunately very difficult for field red giants such as those mentioned above, due to the degeneracy of evolutionary tracks in the HR diagram. Error bars on the stellar mass are often of the order~$\pm 2 M_{\sun}$, making it impossible to determine whether the star under consideration has a mass around 1~$M_{\sun}$ or significantly higher.

To overcome this difficulty we started a high-precision radial velocity survey of red giant stars in a number of intermediate-age open clusters. Knowledge of cluster parameters allows us to accurately determine red giant masses, which we choose between 1.5 and 4~$M_{\sun}$. We obtained first measurements in 2003 with the Coralie spectrograph \citep{queloz00} at the Swiss-1.2m Euler telescope at La Silla Observatory, Chile. However, the faintness of our targets prevented us from observing all of the clusters with Coralie and led us to apply for observing time on a more powerful instrument, the HARPS spectrograph \citep{mayor03} on the ESO-3.6m telescope, also at La Silla. This allowed us to reach a sufficient RV precision on the faintest red giants ($V \cong$ 12) in our sample.

Very recently, following a strategy similar to ours, \citet{sato07} published the discovery of a giant planet orbiting a 2.7~$M_{\sun}$~red giant in the Hyades. This illustrates the potential of open clusters as laboratories to understand planet formation. In this paper we present the discoveries of a massive planet and a brown dwarf around two intermediate-mass stars in our clusters, NGC2423~No3 (TYC~5409-2156-1) and NGC4349~No127 (TYC~8975-2606-1). Sect.~2 describes the sample of stars we are following in our RV survey. The basic properties of both planet-host stars are discussed in Sect.~3, whereas Sect.~4 presents the RV measurements and orbital solutions for both companions. In Sect.~5 we show that the RV variations are not due to spurious stellar effects. Finally, we discuss the implications of these discoveries in the broader context of planet formation in Sect.~6.

\section{Description of the survey}

\begin{table*}
\caption{Cluster list and properties of their red giants.}
\label{TableClusterList}
\centering
\begin{tabular}{c c c c c c c c}
\hline\hline
\bf Cluster & \bf Number of & \bf Cluster age & \bf Mass of giants & \bf Magnitude of \\
 & \bf selected giants & \bf [Gyr] & \bf [$M_{\sun}$] & \bf selected giants \\
\hline
NGC 3114 & 10 & 0.13 $\pm$ 0.05 & 4.7 $\pm$ 0.4 & $\sim$ 8.2 \\
\hline
NGC 4349 & 7 & 0.20 $\pm$ 0.05 & 3.9 $\pm$ 0.3 & $\sim$ 11.3 \\
\hline
IC 2714 & 8 & 0.35 $\pm$ 0.05 & 3.2 $\pm$ 0.2 & $\sim$ 11.2 \\
\hline
NGC 2539 & 9 & 0.37 $\pm$ 0.05 & 3.1 $\pm$ 0.2 & $\sim$ 11.0 \\
\hline
NGC 2447 & 7 & 0.39 $\pm$ 0.05 & 3.0 $\pm$ 0.2 & $\sim$ 10.2 \\
\hline
NGC 6633 & 4 & 0.43 $\pm$ 0.10 & 2.9 $\pm$ 0.2 & $\sim$ 8.7 \\
\hline
IC 4756 & 15 & 0.50 $\pm$ 0.10 & 2.8 $\pm$ 0.2 & $\sim$ 9.2 \\
\hline
NGC 2360 & 8 & 0.56 $\pm$ 0.10 & 2.6 $\pm$ 0.2 & $\sim$ 11.2 \\
\hline
NGC 5822 & 12 & 0.68 $\pm$ 0.20 & 2.5 $\pm$ 0.2 & $\sim$ 10.5 \\
\hline
NGC 2423 & 6 & 0.74 $\pm$ 0.20 & 2.4 $\pm$ 0.2 & $\sim$ 10.5 \\
\hline
IC 4651 & 8 & 1.1 $\pm$ 0.3 & 2.1 $\pm$ 0.2 & $\sim$ 10.8 \\
\hline
NGC 3680 & 4 & 1.2 $\pm$ 0.3 & 2.0 $\pm$ 0.2 & $\sim$ 10.8 \\
\hline
M 67 & 17 & 2.6 $\pm$ 1.0 & 1.5 $\pm$ 0.2 & $\sim$ 10.7 \\
\hline
\end{tabular}
\end{table*}

The clusters in our survey have been selected based on a number of criteria. First of all, to have giant masses between 1.5 and 4~$M_{\sun}$, they have to be of intermediate age (0.2--2~Gyr). They also have to be observable with Coralie and HARPS, so we set a declination limit $\delta \leq$~+15\degr. Moreover, to reach a sufficient photon-limited RV precision ($\leq$10~m~s$^{-1}$), their giants must be brighter than $V=10$ for Coralie and $V=13$ for HARPS. Finally, we only took into account clusters having more than 3~giants known to be non-binary cluster members.

Table~\ref{TableClusterList} gives the list of all selected clusters with their main properties. Cluster ages and data on their red giants have been taken from the WEBDA database \citep{mermilliod95,mermilliod03}. The mass of the red giants in these clusters was derived from their ages using the Padova stellar evolution models at solar metallicity \citep{girardi00}. Error bars on the masses have been estimated taking into account the uncertainties on the cluster ages and metallicities. The assumption of solar metallicity is not likely to cause large errors on the derived masses since the metallicity distribution of open clusters in the solar neighbourhood shows a peak at solar metallicity and a small scatter of $\sim$0.1~dex \citep{twarog97}. The determination of precise cluster metallicities is a challenging task and the values found in the literature for a given cluster are sometimes quite different. This is especially problematic for planet searches since we would like to distinguish between two effects, the well-known planet-metallicity correlation and the influence of stellar mass on planet formation. For this reason we are planning to obtain high signal-to-noise spectra of the giants in our sample in order to perform a high-precision spectroscopic analysis and determine metallicities in the same way as for nearby field stars \citep{santos04b,valenti05,dasilva06}.

\section{Parent star characteristics}

\subsection{NGC2423~No3}

\begin{figure}
\resizebox{\hsize}{!}{\includegraphics{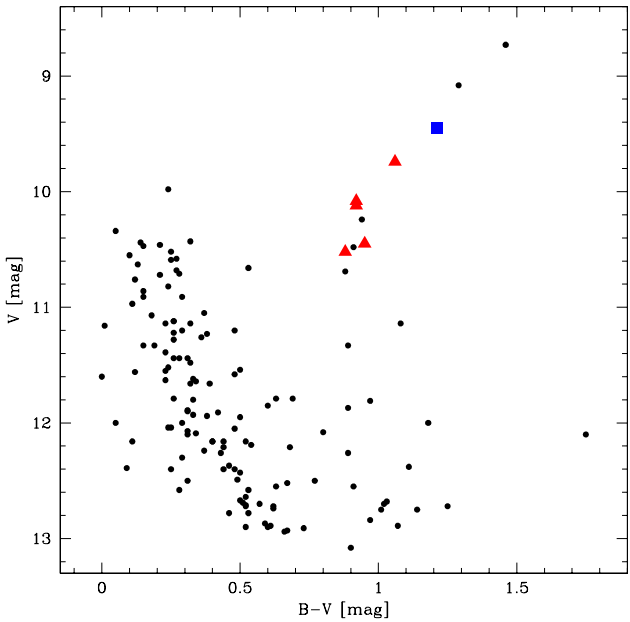}}
\caption{Colour-magnitude diagram for NGC~2423 taken from WEBDA. Red giants in our survey are shown as triangles, NGC2423~No3 as a square.}
\label{FigNGC2423cmd}
\end{figure}

Our sample contains 6 giants belonging to NGC~2423 (distance $d$~=~766~pc). With an age of $\sim$750~Myr, these stars have an estimated mass of $2.4 \pm 0.2 M_{\sun}$. The metallicicty given in WEBDA for this cluster is [Fe/H]~=~0.14~$\pm$~0.09 \citep[from][]{twarog97}. We verified that all 6~stars are indeed cluster members by checking their position in the HR diagram and their radial velocities. Fig.~\ref{FigNGC2423cmd} shows the HR diagram of NGC~2423 as given by WEBDA with the position of our 6~survey stars. All are situated in or close to the red clump. NGC2423~No3 ($\alpha$~=~07h37m09s, $\delta$~=~-13\degr54\arcmin24\arcsec) has visual magnitude $V=9.45$ and colour index $B-V=1.21$ according to \citet{hassan76}. It appears to be somewhat redder and brighter than the mean clump position, indicating that it might be slightly more evolved. However, the small number of red giants in the cluster and the uncertainties in the photometry make it difficult to draw a clear conclusion on its precise evolutionary stage. Regarding radial velocities, we computed from our Coralie measurements a mean radial velocity for the red clump of 18.67~km~s$^{-1}$, with a dispersion of 0.34~km~s$^{-1}$. The mean RV of NGC2423~No3 is 18.32~km~s$^{-1}$, leading to a high probability that it is indeed a cluster member.

\begin{figure}
\resizebox{\hsize}{!}{\includegraphics{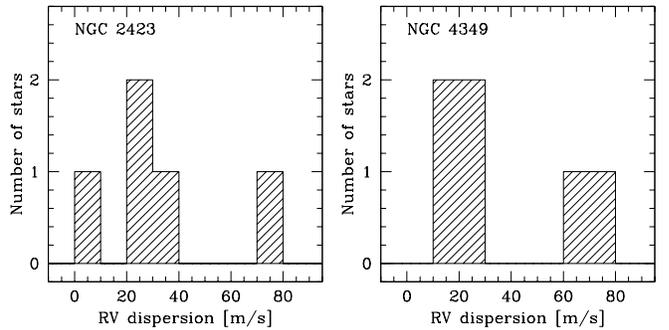}}
\caption{RV dispersion for giant stars in NGC~2423 and NGC~4349. Two stars with dispersions larger than 100~m~s$^{-1}$ (due to stellar companions) are not shown in these plots.}
\label{FigRVdispersion}
\end{figure}

We started the monitoring of the NGC~2423 giants with Coralie in 2003 and accumulated $\sim$10~RV measurements per star, spread over more than 2~years. The average visual magnitude of these stars, $V=10.5$, makes them difficult targets for precision RV measurements with this instrument. Nevertheless we could obtain a typical photon-limited RV precision of $\sim$15~m~s$^{-1}$, sufficient to detect giant planets. Fig.~\ref{FigRVdispersion} shows the histogram of the RV dispersion for the NGC~2423 giants after correction of the instrumental contribution. Four of them exhibit a RV scatter below 30~m~s$^{-1}$, while NGC2423~No3 stands clearly out with its RV dispersion of 72~m~s$^{-1}$. Finally, the last star (NGC2423~No43) shows a large-amplitude, long-term drift ($\sigma_{\mathrm{RV}}$~=~542~m~s$^{-1}$) due to a stellar-mass companion recently discovered by \citet{mermilliod07}.

These results allow us to compute an estimate of the typical RV jitter affecting the NGC~2423 giants. Quadratically subtracting the instrumental contribution, we obtain a value of $\sim$20~m~s$^{-1}$ for the jitter, based on the 4~stable stars in this cluster. It therefore appears that the jitter level will not prevent us from detecting the RV signal of giant planets orbiting these stars.

\subsection{NGC4349~No127}

\begin{figure}
\resizebox{\hsize}{!}{\includegraphics{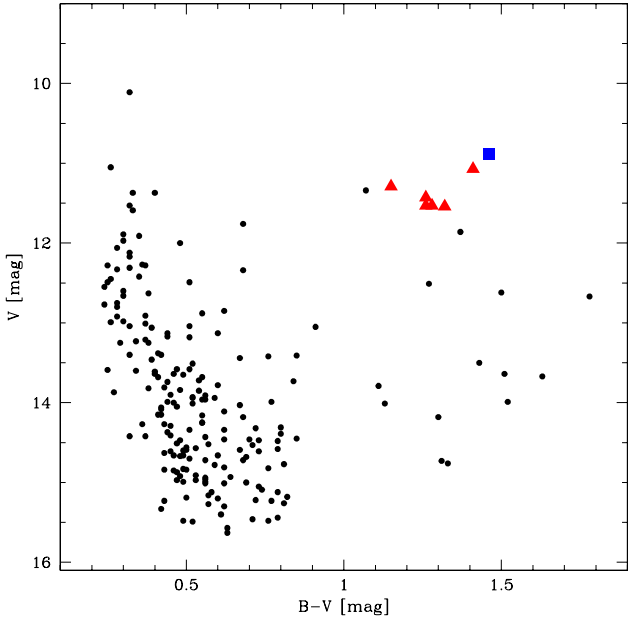}}
\caption{Colour-magnitude diagram for NGC~4349 taken from WEBDA. Red giants in our survey are shown as triangles, NGC4349~No127 as a square.}
\label{FigNGC4349cmd}
\end{figure}

NGC~4349 is a distant open cluster ($d$~=~2200~pc) with an age of only 200~Myr. The best available estimate for the metallicity is [Fe/H]~=~-0.12~$\pm$~0.04 \citep{piatti95}. This is one of the most interesting clusters in our sample since its giant stars have an estimated mass of $3.9 \pm 0.3 M_{\sun}$. We have included 7~of them in our survey, whose positions in the HR diagram are shown in Fig.~\ref{FigNGC4349cmd}. NGC4349~No127 ($\alpha$~=~12h24m35s, $\delta$~=~-61\degr49\arcmin12\arcsec) has visual magnitude $V=10.88$ and colour index $B-V=1.46$ according to \citet{lohmann61}. Again, it seems to be slightly more evolved than the mean clump position. The mean radial velocity of the giants, derived from our measurements, is -11.77~km~s$^{-1}$, with a dispersion of 0.23~km~s$^{-1}$. The mean RV of NGC4349~No127 is -11.40~km~s$^{-1}$; this star is therefore most probably a cluster member.

We use HARPS to observe the NGC~4349 giants since they are too faint for Coralie ($V \cong$ 11.3). To date we have obtained about 7~measurements per star spanning $\sim$500~days. The typical photon-limited RV precision on these stars amounts to 3--4~m~s$^{-1}$. Fig.~\ref{FigRVdispersion} shows the histogram of the RV dispersion after subtraction of the instrumental component. Four stars have a moderate scatter of 13--28~m~s$^{-1}$, while NGC4349~No168 shows a long-term trend ($\sigma_{\mathrm{RV}}$~=~74~m~s$^{-1}$) probably due to a stellar companion. Finally, NGC4349~No203 is a short-period binary exhibiting large RV variations \citep{mermilliod07}. NGC4349~No127 stands out with a scatter of 70~m~s$^{-1}$ and a periodic signal, as will be seen in Sect.~4. Using the stable giants, we derive an estimated jitter level of 20~m~s$^{-1}$, very similar to the NGC~2423 giants. Most interestingly, the high-mass NGC~4349 stars are therefore suitable for planet search with precise radial velocities.

\section{Radial velocity data and orbital solutions}

\subsection{NGC2423~No3}

We obtained a total of 46~data points for NGC2423~No3 (28~with Coralie and 18~with HARPS) over a time span of 1529~days. Tables~\ref{TableRVngc2423coralie} and~\ref{TableRVngc2423harps} give the list of these measurements with their instrumental error bars. These radial velocities have been computed using the standard Coralie and HARPS pipelines. As can be seen, HARPS measurements are of much higher quality than Coralie ones. However, in this case the true uncertainties on the stellar radial velocity are dominated by stellar jitter, which we have estimated to $\sim$20~m~s$^{-1}$ in Sect.~3. In the following we therefore quadratically add this value to the instrumental error to obtain the final error bar. To combine both data sets, we consider HARPS and Coralie as two independent instruments, i.e.
we introduce an RV offset between both instruments as a free parameter when fitting a model to the combined data. We note that the HARPS-Coralie offset cannot be pre-determined once for good, since it depends on the star under consideration, the correlation masks used, etc. In the case of NGC2423~No3, the offset is particularly well constrained since there are time intervals where we have data points from both instruments simultaneously.

\begin{table}
\caption{Radial velocity measurements for NGC2423~No3 obtained with Coralie and their instrumental error bars. All data are relative to the solar system barycenter.}
\label{TableRVngc2423coralie}
\centering
\begin{tabular}{c c c}
\hline\hline
\bf JD-2400000 & \bf RV & \bf Uncertainty \\
 & \bf [km~s$^{-1}$] & \bf [km~s$^{-1}$] \\
\hline
52593.848660 & 18.35282 & 0.00949 \\
52683.598863 & 18.27881 & 0.00849 \\
52693.581623 & 18.26094 & 0.01090 \\
52942.816872 & 18.19154 & 0.01366 \\
53016.761041 & 18.32692 & 0.00918 \\
53045.537306 & 18.32506 & 0.01019 \\
53054.551282 & 18.33671 & 0.00819 \\
53098.521695 & 18.42907 & 0.00835 \\
53110.534244 & 18.42353 & 0.01128 \\
53135.479446 & 18.45580 & 0.01166 \\
53140.475994 & 18.45805 & 0.00814 \\
53288.866103 & 18.38649 & 0.01544 \\
53296.884693 & 18.36009 & 0.00841 \\
53330.852860 & 18.29186 & 0.00849 \\
53363.767436 & 18.23434 & 0.00979 \\
53370.738597 & 18.22681 & 0.00848 \\
53444.592134 & 18.21294 & 0.00762 \\
53449.544167 & 18.20224 & 0.01174 \\
53670.874943 & 18.27125 & 0.01050 \\
53700.849066 & 18.29006 & 0.00711 \\
53703.852968 & 18.29413 & 0.00919 \\
53770.603519 & 18.34491 & 0.01251 \\
53776.661243 & 18.35691 & 0.01013 \\
54005.871155 & 18.36764 & 0.01518 \\
54010.878095 & 18.36103 & 0.00987 \\
54029.847346 & 18.36718 & 0.01191 \\
54031.843317 & 18.33092 & 0.00864 \\
54037.837885 & 18.34742 & 0.00809 \\
\hline
\end{tabular}
\end{table}

\begin{table}
\caption{Radial velocity measurements for NGC2423~No3 obtained with HARPS and their instrumental error bars. All data are relative to the solar system barycenter.}
\label{TableRVngc2423harps}
\centering
\begin{tabular}{c c c}
\hline\hline
\bf JD-2400000 & \bf RV & \bf Uncertainty \\
 & \bf [km~s$^{-1}$] & \bf [km~s$^{-1}$] \\
\hline
53669.846074 & 18.28006 & 0.00142 \\
53674.796790 & 18.27805 & 0.00147 \\
53692.862406 & 18.27738 & 0.00142 \\
53699.842505 & 18.29594 & 0.00140 \\
53721.855948 & 18.31043 & 0.00127 \\
53728.752368 & 18.31082 & 0.00129 \\
53758.656840 & 18.36799 & 0.00189 \\
53764.700939 & 18.34264 & 0.00107 \\
53784.639213 & 18.38299 & 0.00129 \\
53817.563693 & 18.43727 & 0.00142 \\
53831.568277 & 18.41925 & 0.00132 \\
53861.565109 & 18.44917 & 0.00227 \\
54050.825005 & 18.29213 & 0.00148 \\
54054.864340 & 18.31615 & 0.00130 \\
54078.820633 & 18.29799 & 0.00129 \\
54082.782093 & 18.26291 & 0.00183 \\
54114.702923 & 18.26573 & 0.00296 \\
54122.715930 & 18.25682 & 0.00155 \\
\hline
\end{tabular}
\end{table}

\begin{figure}
\resizebox{\hsize}{!}{\includegraphics[angle=270]{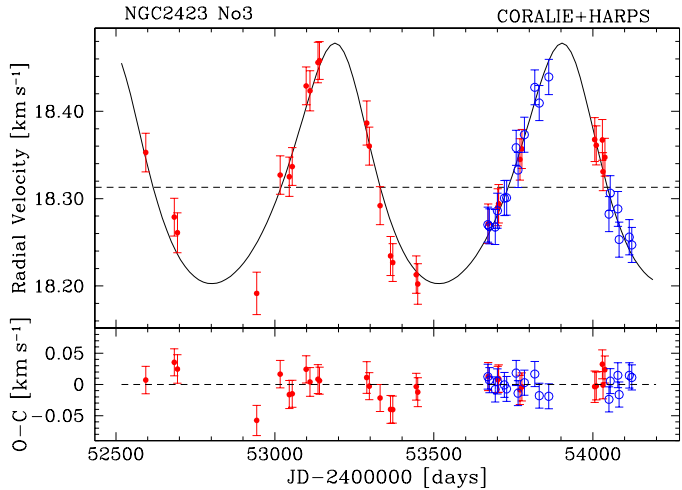}}
\caption{Radial velocity curve as a function of time for NGC2423~No3. The fitted orbit corresponds to a planet with a minimum mass of 10.6~$M_\mathrm{Jup}$ and a period of 714~days. Full dots indicate Coralie data points, while open dots denote HARPS measurements.}
\label{FigNGC2423No3}
\end{figure}

A periodic RV variation is clearly visible in the data, with an approximate period of 700~days (see Fig.~\ref{FigNGC2423No3}). Assuming this signal is due to an orbiting body (see Sect.~5 for a justification), we fit a Keplerian orbit and obtain an orbital period $P$~=~714~days, an eccentricity $e$~=~0.21 and a RV semi-amplitude $K$~=~138~m~s$^{-1}$, leading to a minimum mass $m_2 \sin{i}$~=~10.6~$M_\mathrm{Jup}$ and a semi-major axis $a$~=~2.10~AU for the companion (see Table~\ref{TablePlanets}). The dispersion of the residuals is 18.3~m~s$^{-1}$ and the reduced~$\chi^2$ is 0.84, indicating a good fit to the data, although stellar jitter might have been slightly overestimated.

\subsection{NGC4349~No127}

\begin{figure}
\resizebox{\hsize}{!}{\includegraphics[angle=270]{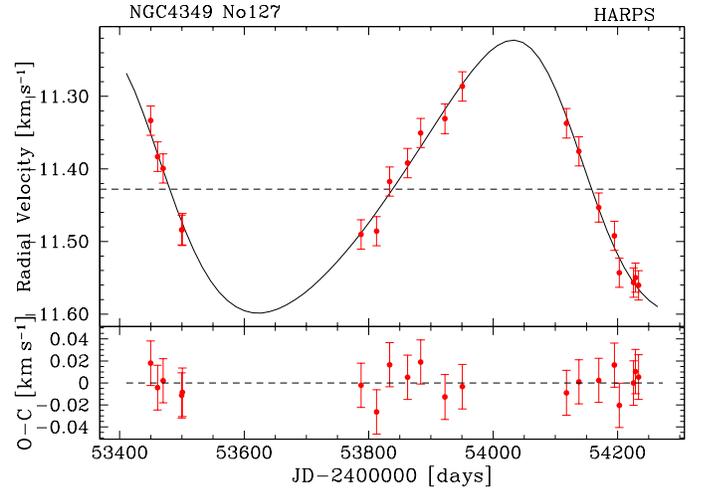}}
\caption{Radial velocity curve as a function of time for NGC4349~No127. The fitted orbit corresponds to a companion with a minimum mass of 19.8~$M_\mathrm{Jup}$ and a period of 678~days.}
\label{FigNGC4349No127}
\end{figure}

We gathered 20 data points for NGC4349~No127 within a time span of 784~days (see Fig.~\ref{FigNGC4349No127}). All these measurements, listed in Table~\ref{TableRVngc4349}, were obtained with HARPS and reduced with the standard high-precision RV pipeline. As for NGC2423~No3, a jitter of 20~m~s$^{-1}$ was quadratically added to the instrumental error bars. Assuming the observed RV variations are due to an orbiting body, we fit a Keplerian orbit to the data, which yields an orbital period $P$~=~678~days, an eccentricity $e$~=~0.19 and a RV semi-amplitude $K$~=~188~m~s$^{-1}$. This corresponds to a minimum mass $m_2 \sin{i}$~=~19.8~$M_\mathrm{Jup}$ and a semi-major axis $a$~=~2.38~AU for the companion (see Table~\ref{TablePlanets}). The RMS of the residuals is 12.6~m~s$^{-1}$ and the reduced~$\chi^2$~0.52. This unreasonably good value probably indicates that stellar jitter was also overestimated in this case. If this RV signal is indeed caused by an orbiting body (as suggested in Sect.~5), then the companion around NGC4349~No127 has a mass above the D-burning threshold and should therefore be referred to as a brown dwarf according to current definitions. Such objects are extremely rare around solar-type stars, and it is therefore remarkable to have detected one among a relatively small sample of intermediate-mass stars (see discussion in Sect.~6).

\begin{table}
\caption{Radial velocity measurements for NGC4349~No127 obtained with HARPS and their instrumental error bars. All data are relative to the solar system barycenter.}
\label{TableRVngc4349}
\centering
\begin{tabular}{c c c}
\hline\hline
\bf JD-2400000 & \bf RV & \bf Uncertainty \\
 & \bf [km~s$^{-1}$] & \bf [km~s$^{-1}$] \\
\hline
53449.782978 & -11.33346 & 0.00246 \\
53460.835689 & -11.38321 & 0.00358 \\
53469.789958 & -11.39941 & 0.00231 \\
53499.577320 & -11.48408 & 0.00444 \\
53500.640853 & -11.48348 & 0.00951 \\
53787.794345 & -11.49031 & 0.00223 \\
53812.758284 & -11.48578 & 0.00274 \\
53833.702527 & -11.41748 & 0.00197 \\
53862.612367 & -11.39202 & 0.00205 \\
53883.589298 & -11.35062 & 0.00174 \\
53922.499118 & -11.33090 & 0.00430 \\
53950.475526 & -11.28637 & 0.00299 \\
54117.845316 & -11.33707 & 0.00404 \\
54137.805538 & -11.37590 & 0.00201 \\
54169.723278 & -11.45319 & 0.00178 \\
54194.778924 & -11.49212 & 0.00226 \\
54202.725180 & -11.54309 & 0.00193 \\
54225.659229 & -11.55665 & 0.00373 \\
54228.668949 & -11.54985 & 0.00266 \\
54233.616624 & -11.56043 & 0.00318 \\
\hline
\end{tabular}
\end{table}

\begin{table*}
\caption{Orbital and physical parameters for the planets/brown dwarfs presented in this paper.}
\label{TablePlanets}
\centering
\begin{tabular}{l l c c}
\hline\hline
\multicolumn{2}{l}{\bf Parameter} & \bf NGC2423~No3 b & \bf NGC4349~No127 b \\
\hline
$P$ & [days] & 714.3 $\pm$ 5.3 & 677.8 $\pm$ 6.2 \\
$T$ & [JD-2400000] & 53213 $\pm$ 21 & 54114 $\pm$ 34 \\
$e$ & & 0.21 $\pm$ 0.07 & 0.19 $\pm$ 0.07 \\
$V$ & [km s$^{-1}$] & 18.3130 $\pm$ 0.0067 & -11.4278 $\pm$ 0.0118 \\
$\omega$ & [deg] & 18 $\pm$ 10 & 61 $\pm$ 19 \\
$K$ & [m s$^{-1}$] & 137.6 $\pm$ 9.1 & 188.0 $\pm$ 13.0 \\
$a_1 \sin{i}$ & [10$^{-3}$ AU] & 8.84 & 11.5 \\
$f(m)$ & [10$^{-6} M_{\sun}$] & 0.180 & 0.442 \\
$m_2 \sin{i}$ & [$M_{\mathrm{Jup}}$] & 10.6 & 19.8 \\
$a$ & [AU] & 2.10 & 2.38 \\
\hline
$N_{\mathrm{meas}}$ & & 46 & 20 \\
\it Span & [days] & 1529 & 784 \\
$\Delta v$ {\tiny (HARPS-Coralie)} & [km s$^{-1}$] & -0.0098 $\pm$ 0.0067 & - \\
$\sigma$ (O-C) & [m s$^{-1}$] & 18.3 & 12.6 \\
\hline
\end{tabular}
\end{table*}

\section{Line shape and activity analysis}

It is always necessary to carefully analyze RV measurements of giant stars since the RV jitter affecting these stars may induce a signal that could be misinterpreted as a planet. The dependence of RV variability on stellar parameters (mass, metallicity, evolutionary stage) is poorly known for giant stars. However, there are indications that clump red giants (i.e. in the core He-burning phase) are intrinsically more stable than first-ascent RGB or AGB stars \citep{bizyaev06}. Short-period (hours to days) and long-term (hundreds of days) RV variations have been known to exist in giant stars for many years \citep[see][]{walker89,larson93,hatzes93,hatzes94}. While the short-period modulations are most probably due to stellar pulsations \citep[e.g. solar-like oscillations, see][]{frandsen02,deridder06,hekker06}, three main reasons may explain the long-term variations: the presence of an orbiting body, rotational modulations of surface features and long-period, non-radial oscillation modes. To distinguish between these hypotheses, we use two well-known diagnostics, the bisector velocity span of the cross-correlation function \citep{hatzes96,queloz01} and the CaII~H\&K~activity index (S-index). The bisector span traces line shape variations and should remain constant if the measured RVs are due to an orbiting body, while the S-index is sensitive to active regions on the stellar surface. In the case of rotational modulations, these indicators should exhibit variations in phase with the radial velocities and the stellar rotation period. The rotation period is however difficult to estimate in our case because relevant parameters such as the stellar radius, the projected rotational velocity, the inclination angle and possible photometric variability are either not known or too uncertain. We therefore rely on the study of bisector and S-index measurements to trace inhomogeneities at the stellar surface. Finally, long-period, non-radial oscillations in G and K giants cannot be excluded but are presently rather hypothetical since they have never been unambiguously identified and lack a clear theoretical support. If present, such oscillation modes should have an effect on the bisector velocity span, although there may be cases where this signature is difficult to measure \citep[see for example][for a detailed discussion]{hatzes99}.

\begin{figure}
\resizebox{\hsize}{!}{\includegraphics{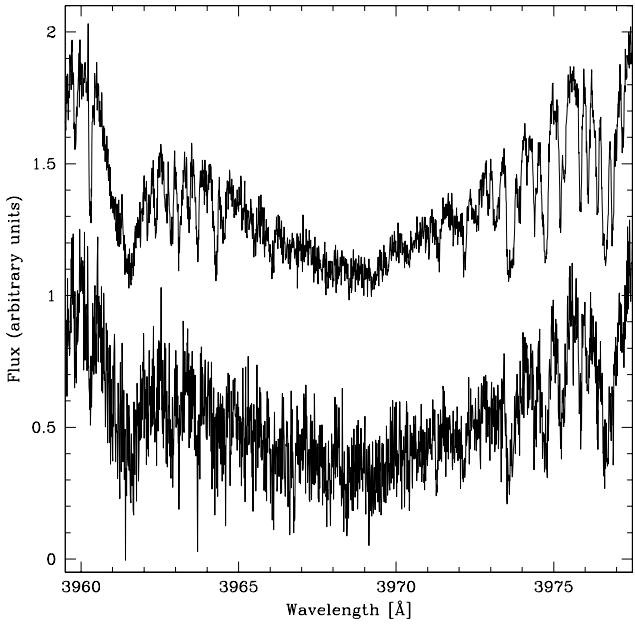}}
\caption{CaII~H spectral region for NGC2423~No3 (top) and NGC4349~No127 (bottom). The spectra have been vertically shifted for clarity. No re-emission features are visible in these giants, indicating a low level of activity.}
\label{FigCaII}
\end{figure}

Fig.~\ref{FigCaII} shows the spectrum of NGC2423~No3 and NGC4349~No127 in the CaII~H region. Although the signal-to-noise ratio is low due to the faintness of the stars, no re-emission features are visible in the CaII~H line core. This suggests a low level of activity in these giant stars. To compute the CaII S-index, we closely follow the procedure originally used at Mount Wilson \citep{vaughan78}, i.e. the fluxes in two narrow bandpasses (1~\AA) centered on the CaII~H\&K cores are integrated and then normalized with the integrated flux in two neighbouring spectral bands. Given the low signal-to-noise ratio, we carefully checked that our measurements and error bars are correct. We made sure that the random errors (photon and detector noise) are properly propagated and checked that no instrumental effects, such as background light pollution, have a significant impact on the computed S-index values. The error bars indicate that we are able to measure CaII flux variations at the 10--20\% level.

\begin{figure}
\resizebox{\hsize}{!}{\includegraphics{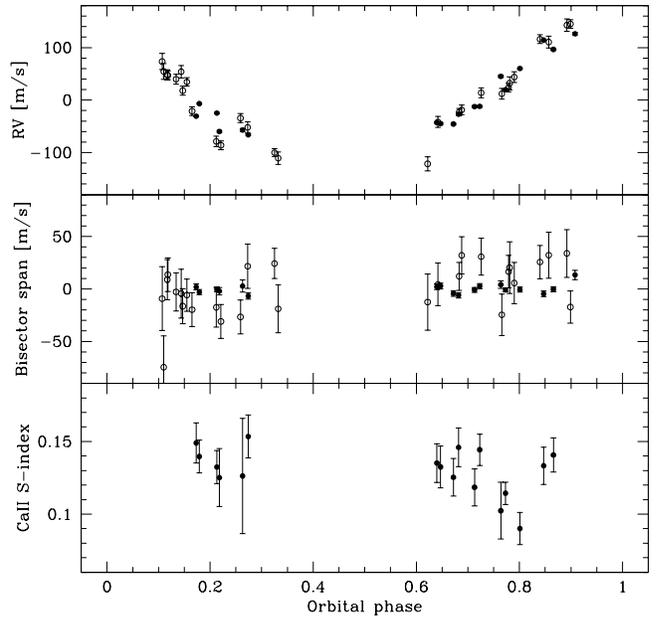}}
\caption{Radial velocity, bisector span and S-index plotted as a function of orbital phase for NGC2423~No3. HARPS and Coralie measurements are shown as full and open dots, respectively. No correlations are seen between the RVs and the other quantities, supporting the planetary hypothesis. Note the different vertical scales for the RV and bisector span measurements.}
\label{FigNGC2423activity}
\end{figure}

\begin{figure}
\resizebox{\hsize}{!}{\includegraphics{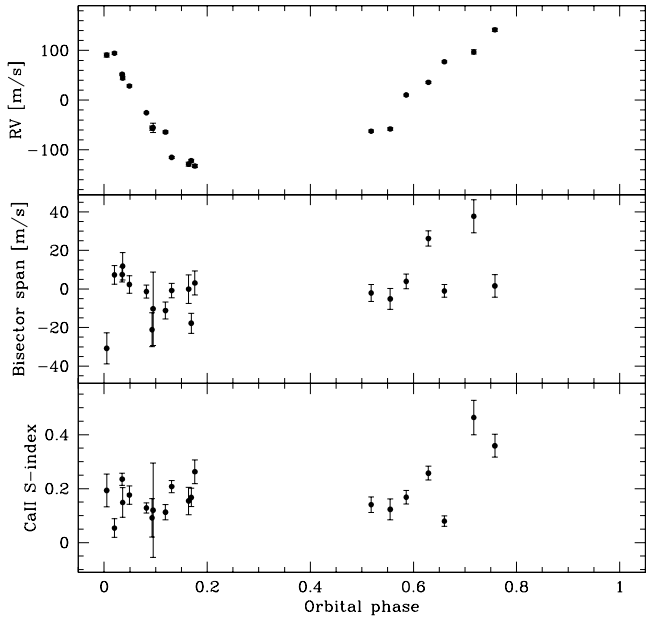}}
\caption{Radial velocity, bisector span and S-index plotted as a function of orbital phase for NGC4349~No127. No correlations are seen between the RVs and the other quantities, supporting the planetary hypothesis. Note the different vertical scales for the RV and bisector span measurements.}
\label{FigNGC4349activity}
\end{figure}

\begin{figure}
\resizebox{\hsize}{!}{\includegraphics{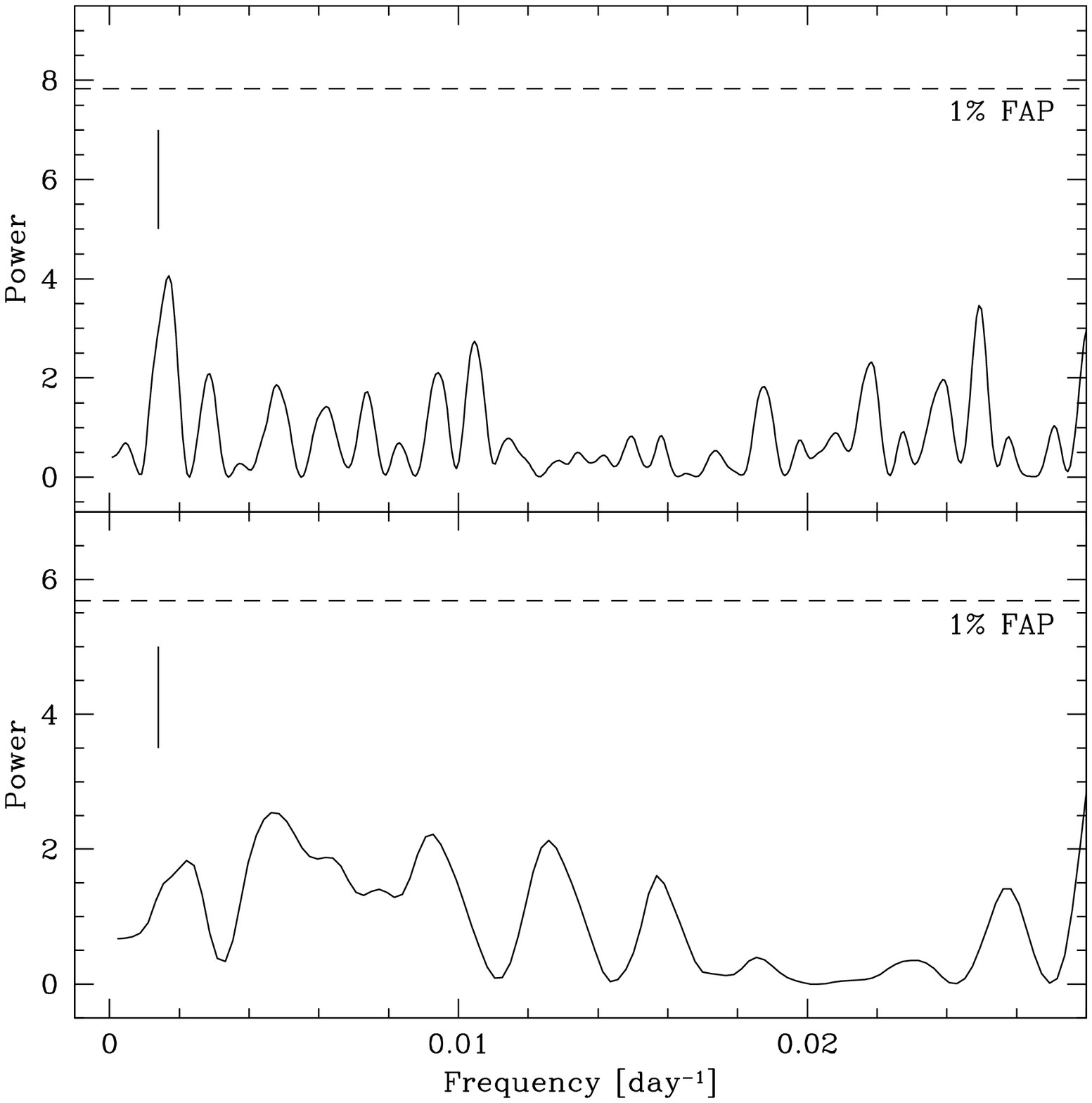}}
\caption{Lomb-Scargle periodograms of the bisector span (top) and S-index (bottom) measurements for NGC2423~No3. The horizontal dashed line indicates the 1\% false-alarm probability threshold. No significant periodicities are found in these quantities. The frequency of the RV signal is shown as a vertical straight line.}
\label{FigNGC2423period}
\end{figure}

\begin{figure}
\resizebox{\hsize}{!}{\includegraphics{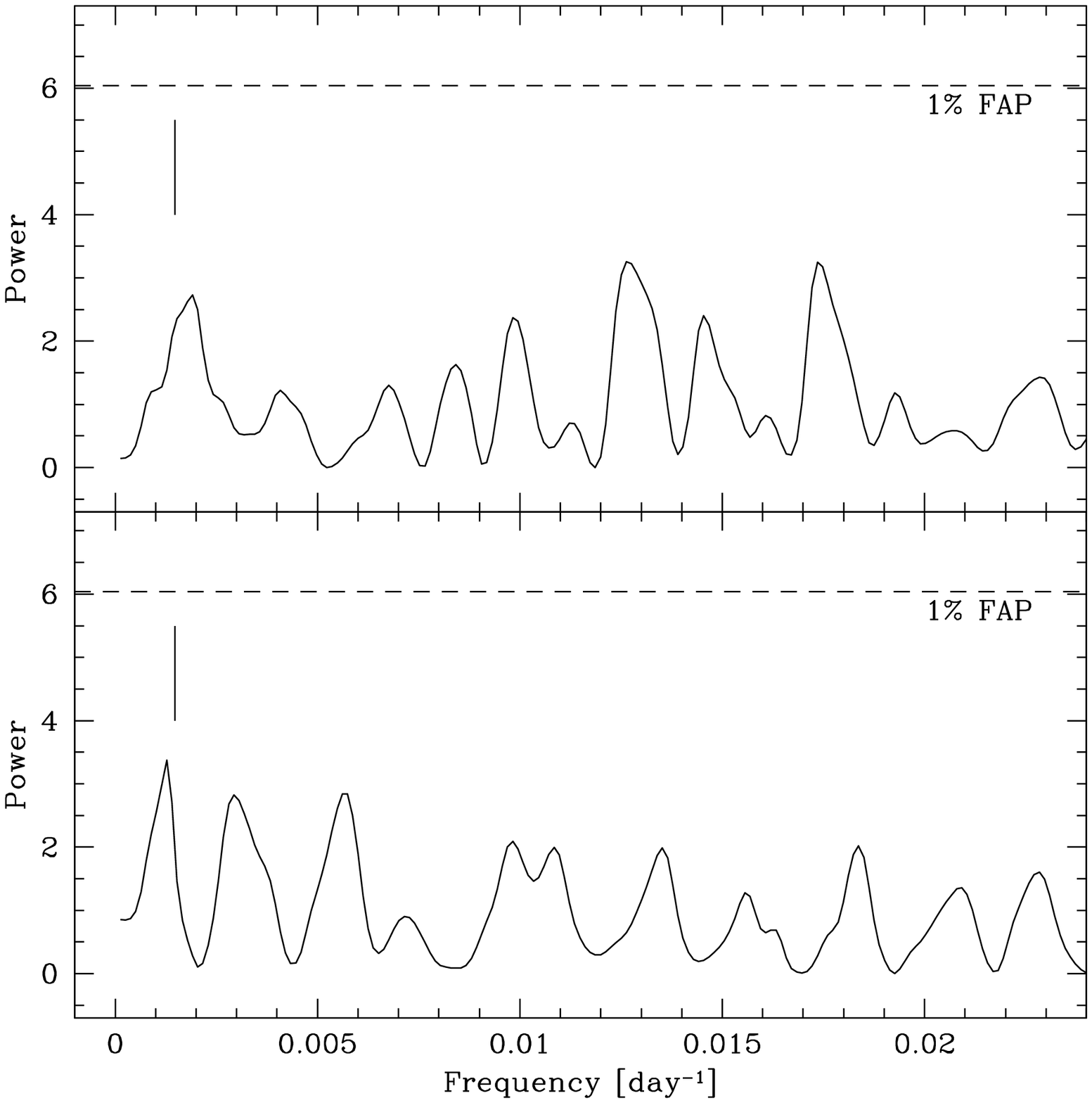}}
\caption{Lomb-Scargle periodograms of the bisector span (top) and S-index (bottom) measurements for NGC4349~No127. The horizontal dashed line indicates the 1\% false-alarm probability threshold. No significant periodicities are found in these quantities. The frequency of the RV signal is shown as a vertical straight line.}
\label{FigNGC4349period}
\end{figure}

Fig.~\ref{FigNGC2423activity} and Fig.~\ref{FigNGC4349activity} show the radial velocity, bisector span and S-index for NGC2423~No3 and NGC4349~No127 as a function of orbital phase. For NGC2423~No3, data from both instruments are shown, except the S-index which could not be computed on Coralie spectra due to the too low signal-to-noise ratio. As can be seen for both stars, the bisector span shows no correlation with the RV variations. HARPS bisector measurements have dispersions of 3.7 and 12.0~m~s$^{-1}$ respectively for NGC2423~No3 and NGC4349~No127, which is about one order of magnitude smaller than the RV variations. Similarly, the S-index does not vary in phase with the RVs (although it shows some variations in the case of NGC4349~No127). To further study the behaviour of the bisector span and the S-index, we computed the Lomb-Scargle periodograms of these two quantities, shown in Fig.~\ref{FigNGC2423period} and Fig.~\ref{FigNGC4349period}. To assess the significance of the peaks, we performed Monte-Carlo simulations to compute their false-alarm probabilities (FAPs). No significant periodicities are found, with the highest peaks always having at least 50\% FAP. The 1\%-FAP level is also indicated in the figures. These results make any further attempts to detect stellar signatures such as rotation or pulsations in these data very speculative. Finally, we also stress that the two stars under consideration are the only ones in their respective clusters showing such a strong RV signal, despite the fact that the other clump giants have similar masses, metallicities and evolutionary status. Altogether, this leads to the conclusion that the planetary hypothesis is the best explanation for the observed, large-amplitude RV variations.

\section{Discussion and conclusion}

\subsection{About planets in open clusters}

Open clusters naturally form homogeneous samples of stars with well-constrained basic properties such as mass, metallicity and age. As a consequence, searching for planets in open cluster environments has the potential to bring a lot of new information on the statistical properties of planetary systems and constrain planet formation and evolution models. As an example, the NGC4349~No127 system is extreme under two aspects: it is the heaviest star (with an accurate mass determination) around which a substellar companion has been detected to date, and it is one of the youngest systems known so far. Its age of 200~Myr represents an upper limit for the timescale of giant planet/brown dwarf formation in the inner regions (a few AUs) surrounding intermediate-mass stars.

With the recent discovery by \citet{sato07}, we now have 3~massive planets or brown dwarfs orbiting intermediate-mass stars in open clusters. This is most probably only the beginning. Our ongoing survey is not complete yet and we plan to extend it to other clusters to increase the significance of statistical analyses. A detailed description of the global results from our survey is beyond the scope of this paper and will be presented elsewhere. Incidentally, we also stress the importance of searching for planets around FGK~dwarfs in open clusters. However, two main problems make such a project difficult: the faintness of the targets (most clusters have FGK~dwarfs fainter than $V=14$), and stellar activity for clusters younger than $\sim$1~Gyr \citep[see for example the RV survey in the Hyades by][]{paulson04}. Nevertheless, future high-precision spectrographs installed on 8--10m telescopes and next-generation ELTs should be able to carry out RV surveys in a large sample of clusters, thereby permitting large-scale comparisons between stars of different masses and metallicities.

There have been a number of studies on the impact of cluster environment on planet formation \citep[see for example][]{armitage00,scally01,bonnell01,smith01,malmberg07}. Close stellar encounters can potentially destroy, or at least strongly affect, planetary systems. Moreover, strong UV radiation from nearby O- and B-stars may have a significant impact on the planet formation process. Little is presently known about the overall magnitude of these effects. Future discoveries of planets in open clusters will help put some constraints on these issues.

\subsection{Planet properties as a function of stellar mass}

Although the sample of intermediate-mass stars being searched for exoplanets is still limited, we can already try to derive some fundamental trends in the characteristics of planetary systems as a function of stellar mass. In particular we would like to compute estimates of the giant planet frequency and typical mass of planetary systems for different categories of stellar masses. In the following we consider three bins of stellar masses, equally spaced on a logarithmic scale: $0.18\leq M/M_{\sun}\leq 0.56$ (mostly M~dwarfs), $0.56\leq M/M_{\sun}\leq 1.78$ (mostly FGK~dwarfs) and $1.78\leq M/M_{\sun}\leq 5.62$ (intermediate-mass red giants). Precise RV surveys targeting the first two bins have been ongoing for more than a decade and have yielded reliable estimates of giant planet frequency around solar-type stars \citep{udry07,marcy06} and M~dwarfs \citep{bonfils06,endl06,butler06}. For the third bin only preliminary results are available, but they seem to show an abnormal number of massive planets or brown dwarfs compared to the other bins, where this kind of objects are extremely rare ("brown dwarf desert"). To quantify this more precisely, we have to define a region in the parameter space of planetary properties where observational biases will not strongly affect the results. In this respect the limiting bin is obviously the category of intermediate-mass stars because of the short duration of the surveys and the increased detection limits due to higher stellar masses and RV jitter. We therefore choose to focus on massive planets with $M>5~M_{\mathrm{Jup}}$ located at orbital distances 0.5~AU~$\leq a\leq$~2.5~AU. Such planets will always induce RV semi-amplitudes larger than 45~m~s$^{-1}$ on periods shorter than 3~years, and should therefore be detectable with $\sim$100\% probability by ongoing RV surveys. We also set a minimal semi-major axis at $a$~=~0.5~AU because intermediate-mass red giants will have engulfed closer-in planets during their evolution along the RGB. Obviously, this is only a rough estimate since the true minimal semi-major axis at which a planet can survive during the RGB phase depends on the detailed stellar structure and system properties.

We now have to estimate for each bin how many stars have been sufficiently observed by RV surveys to reveal massive planets orbiting them. For M~dwarfs, we estimate that about 300~stars meet this criterium if we take into account the high-precision surveys carried out by the HARPS and Elodie teams \citep{bonfils06}, the California-Carnegie team \citep{butler06} and the Texas team \citep{endl06}. For FGK~dwarfs, the total number of stars followed by the different groups over the past decade amounts to about~3000. For intermediate-mass stars, adding our survey to the ones carried out by \citet{sato03} and \citet{setiawan04} leads us to about 200~targets with a mass higher than $\sim$1.8~$M_{\sun}$. Again, these numbers are only rough estimates, but they should nevertheless be correct enough for a qualitative analysis. Other surveys targeting intermediate-mass stars have recently yielded their first results \citep[e.g.][]{johnson07} and should be included in future analyses.

As an input database for exoplanets, we use the Extrasolar Planets Encyclopaedia maintained by J.~Schneider\footnote{http://exoplanet.eu} as of February~2007. We made two modifications to this database. First, we removed the few close binary stellar systems since planets in such systems probably have quite different properties \citep{eggenberger04}. Second, we added to the list a few brown dwarf candidates that were not included since their minimum masses are significantly higher than 13~$M_{\mathrm{Jup}}$. It is necessary to take such objects into account since we want an unbiased census of massive planets and brown dwarfs. However, we had to define an upper mass limit to avoid including objects in the tail of the mass distribution of stellar companions, since we are primarily interested in the distribution of "planetary" companions, to the left of the brown dwarf desert. We put the limit at $M=40~M_{\mathrm{Jup}}$ (in the middle of the desert), which led us to add 3~objects to the database (HD~137510b, HD~180777b and HD~184860b). We note that the exact value of the cut-off mass has no significant impact on the results because of the rarity of brown dwarfs.

We can now compute for each stellar mass bin the frequency of planets and brown dwarfs in the mass and semi-major axis ranges defined above. For M~dwarfs, there are no detected planets meeting the criteria. For solar-type stars, there are 14~planets out of 3000~stars (0.5\%), while for intermediate-mass stars 5~planets have been detected among 200~targets (2.5\%). Assuming the true planet frequency is the same for higher-mass stars as for solar-type stars (i.e.~0.5\%), there is only a 0.3\%~probability to find 5~planets out of 200~intermediate-mass stars. Even if these values are derived from small-number statistics, there seems to be a real trend in the sense that more massive stars form significantly more massive planets or brown dwarfs than lower-mass stars.

\begin{figure}
\resizebox{\hsize}{!}{\includegraphics{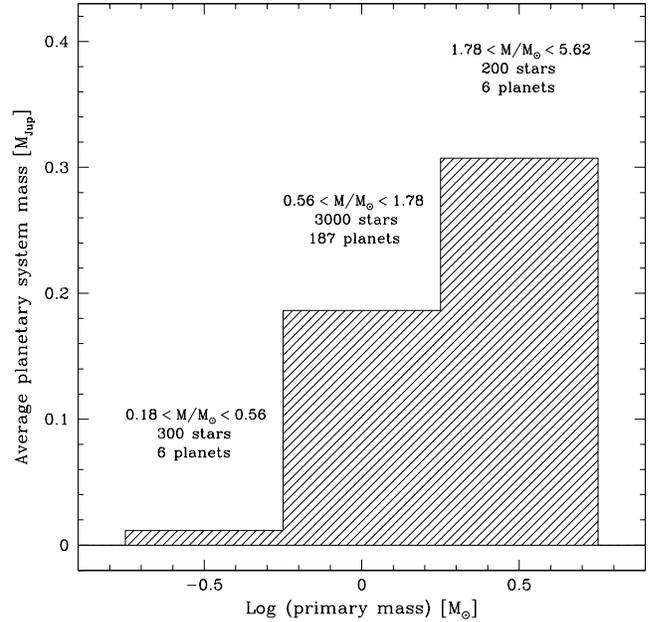}}
\caption{Average mass of planetary systems as a function of stellar mass, taking into account all planets known as of February~2007 (see text for details). More massive stars seem to harbour more massive planetary systems.}
\label{FigAverageMass}
\end{figure}

Alternatively, we can also compute for each bin the average mass of planetary systems, i.e. the total mass of all planets divided by the total number of stars in each bin. We choose this indicator because it gives useful information on the output of the planet formation process (total mass accreted into planetary bodies) and it is relatively free of observational biases. On the one hand, it seems reasonable to assume that most of the mass in planetary systems is contained in the few largest bodies of each system. On the other hand, RV surveys are most sensitive to massive planets. We should then be able to obtain reasonable estimates for the average planetary system mass in the first two bins, while this quantity will probably be underestimated in the third bin due to the short duration and the lower sensitivity of the surveys. Fig.~\ref{FigAverageMass} shows the results in a histogram, which suggests that more massive stars do form more massive planetary systems than lower-mass stars, in spite of the previously mentioned observational biases. If confirmed, this mass scaling raises questions on how to classify objects above 13~$M_{\mathrm{Jup}}$ orbiting solar-type and intermediate-mass stars. An abrupt transition between planets and brown dwarfs would have little meaning if both categories of objects are formed by the same physical process.

Such a scaling in the mass distribution of exoplanets is expected in the core-accretion scenario of planet formation since more massive stars probably have more massive disks, which make it possible to accrete larger amounts of rock, ice and gas. However, more quantitative studies are needed. In the disk-instability paradigm \citep[e.g.][]{boss06}, it is not clear how planet formation depends on stellar mass in general, although \citet{boss06} predicts that this mechanism should not be too sensitive to this parameter. It also remains to be seen if high luminosities and winds will not prevent the formation of gas giants in the inner regions surrounding intermediate-mass stars. As an example, \citet{ida05} predict that the location of the ice boundary at larger distances is likely to make the formation process of gas giants less efficient. They even predict that the fraction of stars harbouring giant planets should decrease beyond 1~$M_{\sun}$, which is in contradiction with the results presented in this paper (at least for planets heavier than 5~$M_{\mathrm{Jup}}$). The apparently high frequency of massive planets around intermediate-mass stars indeed suggests a rather higher efficiency for the accretion process.

\begin{acknowledgements}

We would like to thank the Swiss National Science Foundation (FNRS) for its continuous support. This research has made use of the WEBDA database, operated at the Institute for Astronomy of the University of Vienna.

\end{acknowledgements}

\bibliographystyle{aa}
\bibliography{biblio}

\end{document}